\documentclass[final,1p,times]{elsarticle} 
\usepackage{graphicx}
\usepackage{amssymb} 
\usepackage{amsthm} 
\usepackage{lineno}

\newcommand{\sNN}{{{$\sqrt{s_{_{{NN}}}}$}}}
\newcommand{\dNdeta}{\mbox{$dN_{\mathrm{ch}}/d\eta$}}

\newcommand{\pp}{\mbox{$p+p$}}

\newcommand{\gevc}{\mbox{${\mathrm{GeV/}}c$}}

\newcommand{\muB}{\mbox{$\mu_{B}$}}
\newcommand{\npart}{\mbox{$N_{\mathrm{part}}$}}

\newcommand{\KV}{{\mbox{$\kappa\sigma^{2}$}}}
\newcommand{\SD}{{\mbox{$S\sigma$}}}

\journal{Nuclear Physics A} 

\begin{document}

\begin{frontmatter} 

% Your Title - please insert
\title{Search for the QCD Critical Point by Higher Moments of Net-proton Multiplicity Distributions at STAR}

%% Single author (and collaboration) - please insert
\author{Xiaofeng Luo (for the STAR Collaboration)}
%\fntext[col1] {A list of members of the EMPIRE Collaboration and acknowledgements can be found at the end of this issue.}
\address{ Key Laboratory of Quark$\&$Lepton (MOE) and Institute of Particle Physics, \\ Central China Normal University, 430079 Wuhan, China }
%% Multiple authors
%\author[auth2]{Marcus Junius Brutus}
%\address[auth1]{Somewhere, Rome}
%\address[auth2]{Somewhere else, Rome}

\begin{abstract} 
We present the measurements of the higher moments of the net-proton multiplicity distributions for Au+Au collisions from 
the first phase of the beam energy scan program at RHIC.
The measurements are carried out at midrapidity ($|y|<0.5$) and within the transverse momentum range
$0.4<p_{T}<0.8$ {\gevc} by STAR. The moment products 
, $\KV$ and $\SD$, which are related to the ratios of various order baryon number susceptibilities and sensitive to the correlation length 
, are applied to search for the QCD critical point. We report the centrality and energy dependence of the moment products ({\KV} and {\SD}), results are compared with Poisson expectations. The experiment data are also compared with transport model calculation to understand the non-critical physics effects.
\end{abstract} 

\end{frontmatter} % do not change

%% linenumbers are useful for reviewing process
%\linenumbers

\section{Introduction}
In the year 2010, the Relativistic Heavy Ion Collider (RHIC) launched the Beam Energy Scan (BES) program~\cite{bes} and tuned the Au+Au collision energy from {\sNN}=200 GeV down to 7.7 GeV. The main goal of the BES program is to look for the signature of the first order phase transition and Quantum Chromodynamics (QCD) Critical Point (CP)~\cite{PRL,SQM2011,WWND2011,QM2011_BM}. By varying the colliding energy, we can access a broad region of the QCD phase diagram with a baryon chemical potential ({\muB}) range from about 20 MeV to 450 MeV~\cite{bes}. As the diverges of the correlation length ($\xi$) near the CP, non-monotonic variations of observables with collision energy are believed to be the best signature of CP. Higher moments (skewness ($S$) and kurtosis ($\kappa$) {\it
etc.}) of conserved quantities, such as net-baryon, net-charge and
net-strangeness, distributions are more sensitive to
the correlation length~\cite{qcp_signal,ratioCumulant,Neg_Kurtosis} than the variance ($\sigma^2$) and could be 
detectable if the signature survive in the evolution of the system. On the other hand, the moment products, {\KV} and {\SD}, are also related to the 
ratios of various order susceptibilities as in Lattice QCD~\cite{Lattice,MCheng2009,science}, HRG model~\cite{HRG}
and PQM model~\cite{QM2011_VS}. Ratios of baryon number susceptibilities used to compare with the experimental data as $\kappa
\sigma^2=\chi^{(4)}_{B}$/$\chi^{(2)}_{B}$ and $S
\sigma=\chi^{(3)}_{B}$/$\chi^{(2)}_{B}$ cancel out the volume effect. Theoretical calculations demonstrate that the experimentally measurable net-proton (proton
number minus anti-proton number) number fluctuations can effectively
reflect the fluctuations of the net-baryon number~\cite{Hatta}.
In this paper, we will present measurements for higher moments of event-by-event net-proton
multiplicity distributions from the first phase of the BES program at RHIC.

\section{Results and Discussion}
The data of Au+Au collisions at {\sNN}=7.7, 11.5, 19.6, 27, 39, 62.4 and 200 GeV from the first phase of the BES program 
are used in the analysis. We have also studied {\pp} collisions at {\sNN}=62.4 and 200 GeV. The protons and anti-protons are identified with the ionization energy loss ($dE/dx$) measured by Time Projection Chamber (TPC) at midrapiidty ($|y|<0.5$) and within the transverse momentum range $0.4<p_{T}<0.8$ {\gevc}. To suppress autocorrelation effects between measured net-proton fluctuations and centrality in our results, centralities are determined by the uncorrected charge particle multiplicity ({\dNdeta}) by excluding the protons and anti-protons within pseudorapdity $|\eta|<0.5$, a new technique add to the moment analysis after QM2011 conference.
If protons and anti-protons are independent Poissonian distributions, the net-proton multiplicity will follow the skellam
distribution, which is expressed as:  $$P(N) = {(\frac{{{M_p}}}{{{M_{\overline p}}}})^{N/2}}{I_N}(2\sqrt {{M_p}{M_{\overline p}}} )\exp [ - ({M_p} + {M_{\overline p}})],$$ where $I_{N}(x)$ is a modified Bessel function, $M_{p}$ and $M_{\overline p}$ are the measured mean values of protons and anti-protons. The various order cumulants ($C_{n}$) are closely connected with the moments, e.g., $C_{1}=<N>=M,C_{2}=<(\Delta N)^2>=\sigma^{2}, C_{3}=<(\Delta N)^3>=S \sigma^{3}, C_{4}=<(\Delta N)^4>-3<(\Delta N)^2>^2=\kappa \sigma^{4}$. If the net-proton follows the skellam distribution, then we have,  $S\sigma  = {C_3}/{C_2} = ({M_p} - {M_{\overline p }})/({M_p} + {M_{\overline p }})$ and $\kappa {\sigma ^2} = {C_4}/{C_2} = 1$, which provide the Poisson expectations for the moment products.
\begin{figure}[htb]
 \hspace{-1.8cm}
\begin{minipage}[t]{0.6\linewidth}
\centering \vspace{0pt}
    \includegraphics[scale=0.4]{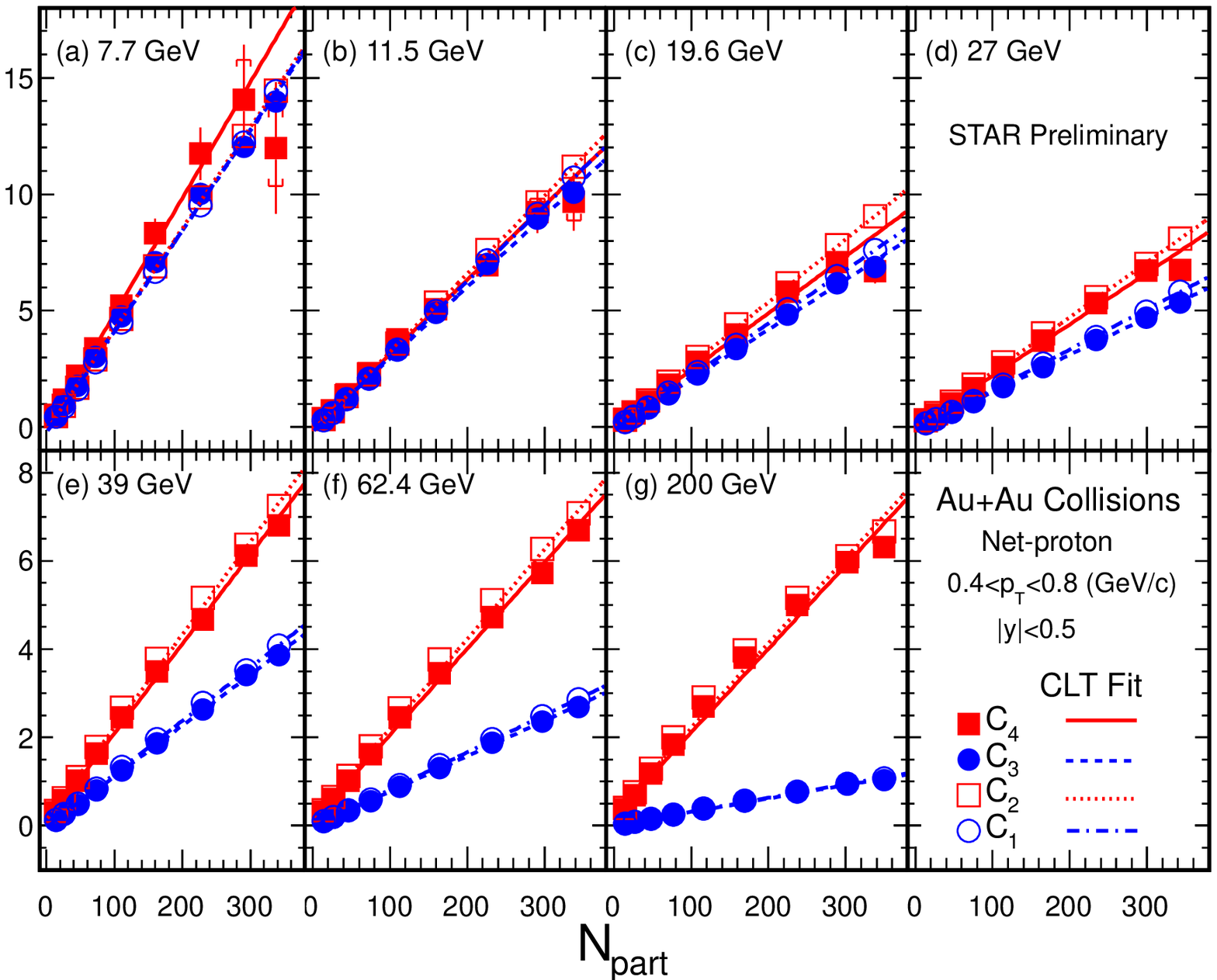}
   \caption{(Color Online) Centrality dependence of the cumulants of net-proton distributions
for Au+Au collisions at {\sNN}=7.7, 11.5, 19.6, 27, 39, 62.4 and 200 GeV. The lines are linear fit to these cumulants. Errors bars are statistical and caps are systematic errors.}
\label{fig:CLT}
  \end{minipage}%
  \hspace{0.2in}
  \begin{minipage}[t]{0.6\linewidth}
  \centering \vspace{0pt}
   \includegraphics[scale=0.4]{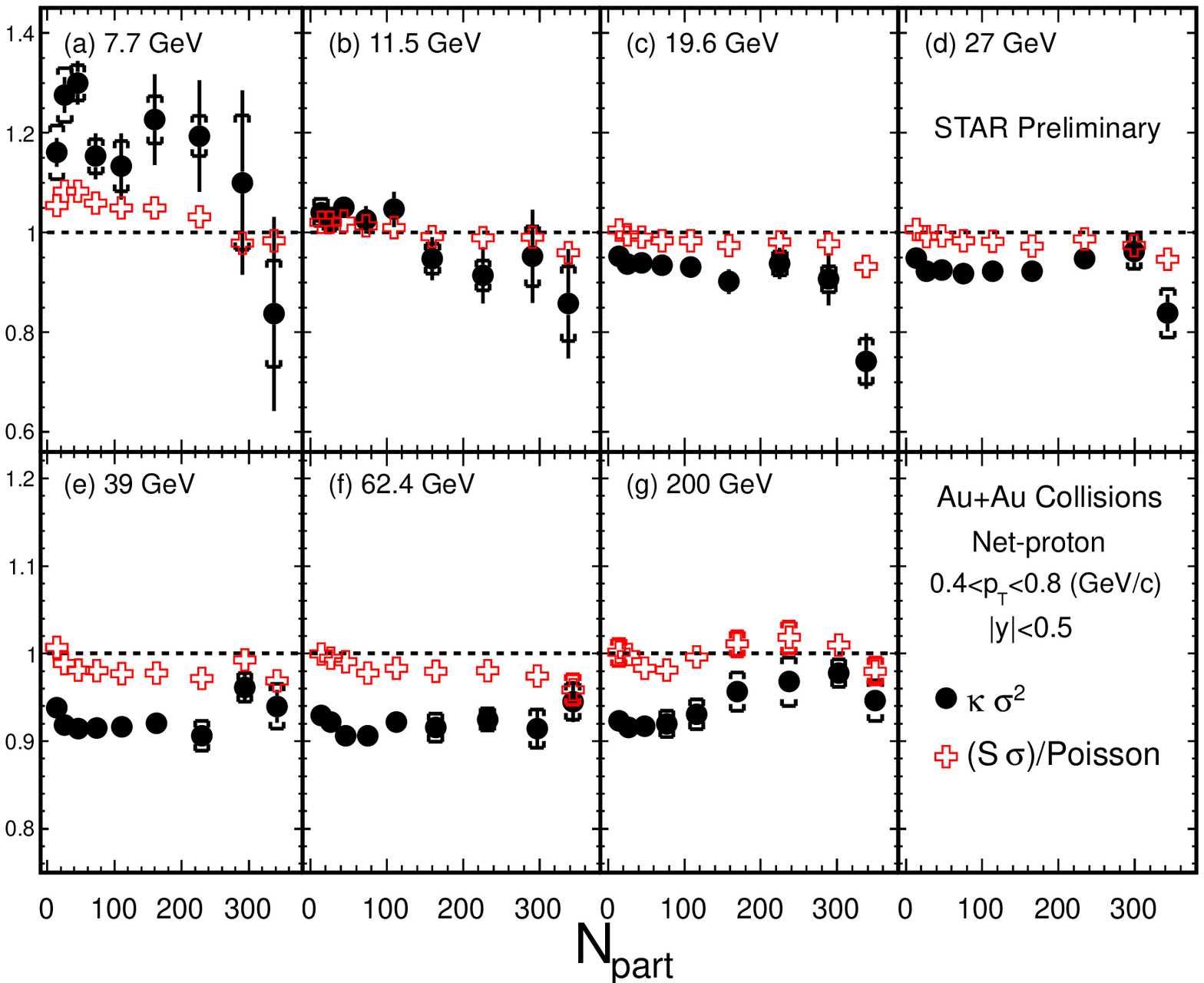}
    \caption{(Color Online) Centrality dependence of {\SD}/Poisson and {\KV} of net-proton
    distributions for Au+Au collisions at {\sNN}=7.7, 11.5, 19.6, 27, 39, 62.4 and 200 GeV.
    The error bars are statistical and caps are systematic errors.} \label{fig:SD_KV_Centrality}
  \end{minipage} %
\end{figure}

The various order cumulants ($C_{1}-C_{4}$) shown in Fig. \ref{fig:CLT} are obtained from the net-proton distributions and correct for the finite centrality bin width effect~\cite{WWND2011}. 
The statistical error estimation in Fig.\ref{fig:CLT} are based on the Delta theorem method~\cite{Delta_theory}, which
is an updated method compare to the analytical method used in QM2011 results~\cite{QM2011_BM}. The systematical errors are estimated by varying the track quality cut condition and particle identification. $C_{1}-C_{4}$ values in Fig.\ref{fig:CLT} show linear increase with {\npart} in each centrality. The lines shown in the Fig.\ref{fig:CLT} are linear fit to those cumulants, which is motivated by the Central Limit Theorem (CLT). In Fig. \ref{fig:CLT}, it is observed that the odd ($C_{1}$ and $C_{3}$) and even ($C_{2}$ and $C_{4}$) order cumulants group together.
The difference between the two groups are smaller at lower energies. The ratios of the cumulants, which are connected to the moment products as $S\sigma  = {C_3}/{C_2}$ and $\kappa {\sigma ^2} = {C_4}/{C_2}$, are shown in Fig. \ref{fig:SD_KV_Centrality}. It is observed that the {\KV} and the {\SD} values normalized to Poisson expectations are below unity for {\sNN} above 11.5 GeV and above unity for 7.7 GeV in Au+Au collisions. The data presented here also provide information to extract freeze-out conditions in heavy-ion collisions using QCD based approaches~\cite{CPOD2011_FKarsch}. 
\begin{figure}[htb]
\begin{center}
 \includegraphics[width=0.65\textwidth]{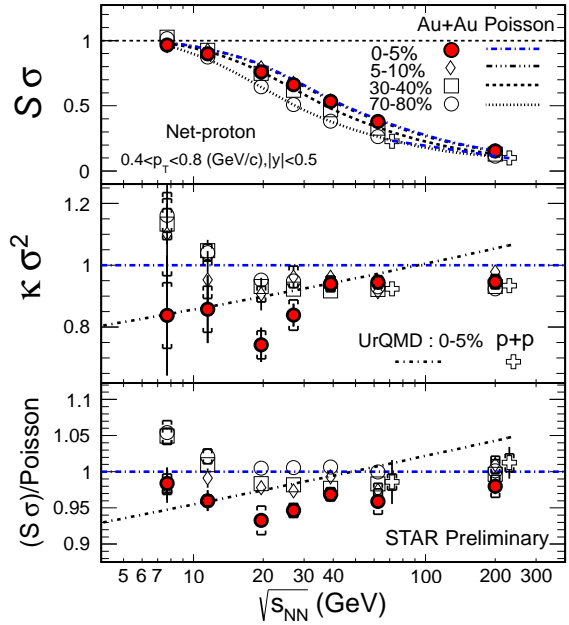}
\end{center}
\vspace{-0.3cm}
\caption{(Color online) Energy dependence of {\KV} and {\SD} for net-proton distributions for four
collision centralities (0-5\%, 5-10\%, 30-40\% and 70-80\%) measured at STAR. The results are compared 
to UrQMD model calculations and {\pp} collisions at {\sNN}=62.4 and 200 GeV. The lines in top panel are the Poisson
expectations and in the bottom panel shows the {\SD} normalized to the corresponding Poisson expectations. } 
\label{fig:SD_KV_Energy}
\end{figure}

Fig.\ref{fig:SD_KV_Energy} shows the energy dependence of {\KV} and {\SD} of net-proton distributions for four centralities (0-5\%, 5-10\%, 30-40\% and 70-80\%) in Au+Au collisions. Bottom panel of Fig.\ref{fig:SD_KV_Energy} shows {\SD} values normalized to the corresponding Poisson expectations. The {\KV} and normalized {\SD} values are similar for all of the centralities in Au+Au collisions at {\sNN}=39, 62.4 and 200 GeV and differences are observed 
between the 0-5\% central Au+Au collisions and the peripheral collisions below {\sNN}=39 GeV. The peripheral collision results are closer to unity and for {\sNN}=7.7 GeV, those are above unity. The UrQMD model~\cite{urqmd} results are also shown in the 
Fig. \ref{fig:SD_KV_Energy} for 0-5\% centrality to understand the non-CP effects, such as baryon number conservation 
and hadronic scattering. The UrQMD calculations show a monotonic decrease with decreasing beam energy.

\section{Summary}
We present the beam energy ({\sNN}=7.7$-$200 GeV) and centrality dependence for the higher moments of net-proton distributions in Au+Au collisions from the first phase of the 
BES program at RHIC. We have discussed the application of those observables to search for the QCD critical point. 
It is observed that the moment products, {\KV} and {\SD}, are below the Poisson expectation for central Au+Au collisions. The peripheral collision results are closer to unity and for {\sNN}=7.7 GeV, those are above unity. The experimental data are also compared with the UrQMD model calculation to understand the non-critical physics effects. The UrQMD calculations show a monotonic decrease with decreasing beam energy. Due to the small correlation length and dynamical evolution of the system in the heavy ion collisions, the signature of the CP could be small. This needs comparison between our data and QCD calculations with dynamical evolution of the system. We also need more statistics to get precise measurements below 19.6 GeV and additional data at {\sNN}=15 GeV. These are planned for the second phase of the BES program at RHIC.

\section*{Acknowledgement}
The work was supported in part by the National Natural Science
Foundation of China under grant No. 11205067 and 11135011. CCNU-QLPL Innovation Fund(QLPL2011P01)
and China Postdoctoral Science Foundation (2012M511237).
\section*{References}
\bibliography{QM2012}

\begin{thebibliography}{10}

\bibitem{bes}
M. M. Aggarwal {\it et al.} (STAR Collaboration), arXiv: 1007.2613.

\bibitem{PRL}
M. M. Aggarwal {\it et al.} (STAR Collaboration), \PRL {\bf 105}, 022302
  (2010).

\bibitem{SQM2011}
X. Luo (for the STAR Collaboration), Acta Phys. Pol. B Proc. Supp.{\bf 5}, 497
  (2012). [arXiv:1111.5671].

\bibitem{WWND2011}
X. Luo (for the STAR Collaboration), {J. Phys.: Conf. Ser.} {\bf 316}, 012003
  (2011).

\bibitem{QM2011_BM}
B. Mohanty (for the STAR Collaboration), J. Phys. G: Nucl. Part. Phys. {\bf
  38}, 124023 (2011).

\bibitem{qcp_signal}
M. A. Stephanov, \PRL {\bf 102}, 032301 (2009).

\bibitem{ratioCumulant}
C. Athanasiou {\it et al.}, \PRD {\bf 82}, 074008 (2010).

\bibitem{Neg_Kurtosis}
M. A. Stephanov, \PRL {\bf 107}, 052301 (2011).

\bibitem{Lattice}
R. V. Gavai and S. Gupta, \PLB {\bf 696}, 459 (2011).

\bibitem{MCheng2009}
M. Cheng {\em et al.}, { Phys. Rev. D} {\bf 79}, 074505 (2009). A. Bazavov,
  {\it et al.}, arXiv:1203.0784.

\bibitem{science}
S. Gupta, X. Luo, B. Mohanty, H. G. Ritter, N. Xu, {Science} {\bf 332}, 1525
  (2011).

\bibitem{HRG}
F. Karsch and K. Redlich, \PLB {\bf 695}, 136 (2011).

\bibitem{QM2011_VS}
V. Skokov {\it et al.}, J. Phys. G: Nucl. Part. Phys. {\bf 38}, 124102 (2011).

\bibitem{Hatta}
Y. Hatta and M. A. Stephanov, \PRL {\bf 91}, 102003 (2003).

\bibitem{Delta_theory}
X. Luo, J. Phys. G: Nucl. Part. Phys. {\bf 39}, 025008 (2012).

\bibitem{CPOD2011_FKarsch}
F. Karsch, CPOD2011 proceedings, arXiv:1202.4173; A. Bazavov, {\it et al.},
  arXiv:1208.1220.

\bibitem{urqmd}
M. Bleicher {\it et al.}, J. Phys. G: Nucl. Part. Phys. {\bf 25}, 1859 (1999).

\end{thebibliography}
\bibliographystyle{unsrt}
\end{document}